\documentclass{ws-procs9x6}

\usepackage{color}

\newcommand{\pa}{\partial}
\begin{document}

\title{CHAPTER X\\~\\Casimir Effect Associated with Fractional \\
   Klein-Gordon Field
 }

\author{S. C. Lim}

\address{Faculty of Engineering,
Multimedia University,\\  
Cyberjaya, 63100,
Selangor, Malaysia}

\author{L. P. Teo}

\address{  Faculty of Engineering, University of Nottingham Malaysia Campus,\\
Semenyih, 43500, Selangor, Malaysia }

\begin{abstract}
This paper gives a brief review on the recent work on fractional Klein-Gordon fields, in particular on the Casimir effect associated to fractional Klein-Gordon fields in various geometries and boundary conditions. New results on Casimir piston due to a fractional Klein-Gordon massive field are given.
\end{abstract}

\bodymatter

\section{Introduction}\label{sec1}
 The possibility of the fractal structure of spacetime has gained ground lately due to the work on quantum theory of gravity. One promising theory known as causal dynamical triangulations proposed by Ambj{\o}rn, Jurkiewicz and Loll \cite{1,2,3} is a non-perturbative approach to quantum gravity which shows how the spacetime has fractal structure, with the number of dimensions changes smoothly from the usual four at large scales to two at the Planck scale. Another approach of quantum gravity, the asymptotically safe quantum Einstein gravity \cite{4,5} also has a similar property such that spacetime evolves from a lower dimensional scale-invariant structure at high energy or micro scales to that of normal four-dimensional geometry at large scales or low energy. In these theories, spacetime geometry cannot be understood in terms of a single metric. There is a need to introduce a different effective metric at each energy scale. Modesto  \cite{6} has applied a similar technique used in asymptotically safe quantum gravity to loop quantum gravity and obtained a similar fractal-like spacetime structure. Numerical simulations of the propagation of a scalar particle in a dynamically triangulated spacetime, with a discretized Einstein dynamics have been carried out. It is found that the spectral dimension of the microscopic spacetime is two and it becomes four for large scales \cite{4,7,8,9,10}~. Another proposal of a Lorentz invariant theory in a fractal spacetime requires the spectral dimension and the ultraviolet Hausdorff dimension of the spacetime to be both equal to two \cite{11}~. The fractal nature of quantum spacetime may provide a way out of the ultraviolet problem with the reduction of spacetime dimensions. It is interesting to note that in addition to these   different approaches to quantum gravity which indicate the emergence of a ground-scale spacetime with fractal properties, there also exists a recent work \cite{12} which shows that spacetime with quantum group symmetry in general has scale-dependent dimensions, and it allows one to establish a link between non-commutative geometries and quantum group and seemingly unrelated approaches of quantum gravity like causal dynamical triangulations and quantum Einstein gravity. All these different theories have a common property, namely, dimensions of spacetime depend on scale, and they could have fractal properties at small scales. At a more fundamental level, Palmer \cite{13} introduced the postulate of an invariant set which assumes the existence of a state space embedding an invariant fractal subset which represents physical reality. It provides a possible link between fractal geometry and quantum theory and it may help to resolve some basic problems in quantum mechanics.

One useful mathematical tool in dealing with fractals and fractal-related phenomena is fractional calculus \cite{14,15,16,17,18}~. An important development that stimulates the application of fractional calculus to physical sciences is the realization of the close connection between fractional calculus and fractal geometries \cite{19,20,21,22}~. Such a relation makes fractional differential equations a natural tool for describing transport processes with fractal properties, such as anomalous diffusion, non-Debye relaxation processes, and other fractal phenomena \cite{16,17,23,24}~. Although the application of fractional differential equations to model various transport phenomena in complex heterogeneous media becomes more and more popular during the past two decades, their use in quantum theory is still quite limited. For many years the trajectories of the well-known fractal process, the Brownian motion, have been extensively used in Feynman path integral approach to quantum mechanics \cite{25}~. The paths of the Brownian motion are known to be fractal curves with Hausdorff dimension two \cite{26}~. Brownian motion also plays an important role in stochastic mechanics, which provides an alternative formulation to quantum mechanics \cite{27}~. In the early application of fractal geometry in quantum field theory, the main focus was on the quantum field models in fractal sets, and quantum field theory of spin systems such as Ising spin model \cite{28,29}~. Fractal geometry was later applied to Wilson loops in lattice gauge theory, and in the modeling of random surfaces in quantum gravity \cite{30}~.

The possible existence of fractal structure of quantum spacetime prompted one to employ fractional calculus and fractional differential equations \cite{14,15,16,17,18}in quantum theory. Just like fractional diffusion equation has replaced diffusion equation to describe anomalous transport systems in fractal media, for quantum theories in fractal spacetime it is natural to consider quantum mechanics and quantum field theory which satisfy fractional generalizations of Schr\"odinger equation, Klein-Gordon equation and Dirac equation. There are some studies on the formulation of quantum mechanics based on various types of fractional Schr\"odinger equations and their possible applications \cite{31,32,33,34,35,36,37,38}~. By noting that fractional diffusion equations are just the Euclidean Schr\"odinger equations, results obtained for the former can be directly applied to the latter. In particular, fractional L$\acute{\text{e}}$vy path integral approach of fractional quantum mechanics has been considered  \cite{31,32,39,40}~. It will be interesting to see whether it is possible to use the path integral representation of fractional Brownian motion \cite{41,42} and fractional oscillator processes \cite{43} in the formulation of fractional quantum theory. It has been shown that fractional Brownian motion does appear naturally in the fractional gluon propagator with the temporal gauge condition \cite{44}~. Incorporation of supersymmetry into fractional quantum mechanics has also been considered \cite{45}~.

Fractional Klein-Gordon equation and fractional Dirac equation have been studied by several authors during the past decades \cite{46,47,48}~. Fractional power of D$^{\prime}$Alembertian operator was used in the nonlocal kinetic terms Lagrangian field theory in the (2+1)-dimensional bosonization. Such fractional power operators also appeared in the effective field theory which has some degrees of freedom integrated out from the underlying local theory \cite{49,50,51}~. Canonical quantization of fractional Klein-Gordon massless and massive fields has been studied by some authors \cite{52,53}~. Quantization of fractional Klein-Gordon field and fractional gauge field based on Nelson's stochastic mechanics and Parisi-Wu stochastic quantization procedure at zero and positive temperature have been considered \cite{44,54}~. There also exists work on axiomatic approach to fractional Klein-Gordon field, where properties of the $n$-point Schwinger or Euclidean Green functions and their analytic continuation to the corresponding $n$-point Wightman functions were given \cite{55,56}~. Fractal fermion propagator has been obtained as a consequence of the QED radiative corrections, which requires the incorporation into the propagator with a fractional exponent connected with the fine structure constant \cite{57}~. More recent work is related to the Casimir effect for the massless and massive fractional fields at zero and positive temperature; and work on Casimir effect due to fractional Klein-Gordon field subject to fractional Neumann boundary conditions which interpolate between the usual Dirichlet and Neumann conditions have been carried out \cite{58}~.

In this paper, we shall mainly concern with the Casimir effect associated with fractional Klein-Gordon fields. In the next section a brief review on the results obtained for the Casimir energy and Casimir force due to a fractional Klein-Gordon field will be given. We shall also discuss the implication of fractional Neumann boundary conditions on the Casimir force. Calculations of Casimir force related to piston geometry for the massive fractional Klein-Gordon field will be given in detail in Section \ref{sec6}~. In the concluding section some possible directions for further work will be discussed.

\section{Casimir Effect Associated with Fractional Klein-Gordon Field}\label{sec3}

Currently, the biggest challenge in cosmology is the explanation of the accelerated expansion of the universe. This is closely related to the problem of dark energy which has been proposed to explain the cause of this accelerating expansion. Various candidates of dark energy have been considered, among them are positive cosmological constant, quintessence, Casimir energy in extra spacetime dimensions, etc \cite{59,60}~.  As a possible candidate for dark energy, Casimir energy \cite{63,64,65,66}~ has attracted considerable renewed interest. On the other hand, Casimir effect has become significant in nanotechnology. At the nano-scales, the Casimir force cannot be neglected, and therefore scientists working in nanotechnology begin to show interest in this subject. Improvement in the ability to measure small forces near surfaces in nano-devices has allowed the possibility to manipulate Casimir force to drive a nano-device. Experimental advances together with improved computations and simulations of Casimir forces make it possible to tailor a nano-device to eliminate unwanted properties caused by the Casimir force such as stiction or adhesion, and hence facilitate the design of better and more effective nano-devices. Thus, from both large and small scales Casimir effect seems to play an important role, as it is expected from vacuum energy which is all prevalent in the universe \cite{62}~.

\begin{figure}\centering \epsfxsize=.8\linewidth
\epsffile{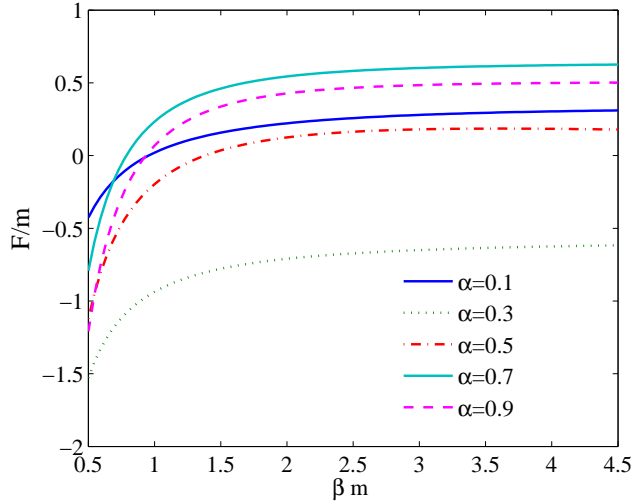}\caption{ \label{f2}
$\mathcal{F} /m $ as a function of
$\beta m $ when $\alpha=0.1, 0.3, 0.5, 0.7, 0.9$ and
$\gamma=1$}\end{figure}

\begin{figure}\centering \epsfxsize=.8\linewidth
\epsffile{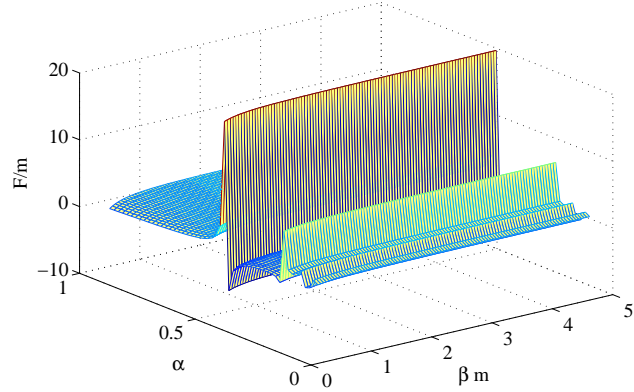}\caption{\label{f1}
$\mathcal{F}/m$ as a function of
$\beta m$ and $\alpha$. Here $\gamma=1$.}\end{figure}

In this section we summarize results on some physical quantities related to Casimir effect due to fractional Klein-Gordon fields. Denote by $\displaystyle \Delta=\frac{\pa^2}{\pa t^2}+\sum_{j}^D \frac{\partial^2}{\partial x_j^2}$  the $(D+1)$-dimensional Euclidean Laplacian operator. The Riesz fractional derivative  $(-\Delta)^{\alpha}$ is defined as \cite{14,18}:
\begin{equation*}
(-\Delta)^{\alpha}f(t,x)=F^{-1}\left((k_0^2+k^2)^{2\alpha}\hat{f}(k_0,k)\right),
\end{equation*}
where $\hat{f}(k_0,k)=F(f(t,x))$   is the Fourier transform of  $f(t,x)$. Let  $\phi(t,x)$, $t\in\mathbb{R}$, $x\in \mathbb{R}^D$,  be the real scalar Klein-Gordon field. Consider the Lagrangian density
\begin{equation*}
L=-\frac{1}{2}\phi(t,x)\Lambda(\pa)\phi(t,x),
\end{equation*}
where $\Lambda(\pa)$  is a pseudo-differential operator. The resulting Lagrangian equation is then given by $\Lambda(\pa)\phi(t,x)=0$. For the scalar massive fractional Klein-Gordon field, one can have the following possible choices of fractional Klein-Gordon operator:
\begin{align*}
\Lambda(\pa)=&(-\Delta+m^2)^{\gamma}\hspace{7cm}\text{(I)}\\
\Lambda(\pa)=&(-\Delta)^{\alpha}+m^{2\alpha}\hspace{6.75cm}\text{(II)}\\
\Lambda(\pa)=&\left[(-\Delta)^{\alpha}+m^{2\alpha}\right]^{\gamma}\hspace{6.1cm}\text{(III)}
\end{align*}
Here we remark that I seems to be the simplest choice, and II is a special case of III when  $\gamma=1$.
Simple results on Casimir effect associated with fractional Klein-Gordon field of type I have been obtained\cite{54}~. Note that the zero temperature fractional Klein-Gordon scalar massive field with Euclidean propagator is given by $S(k)=(k^2+m^2)^{-\gamma}$; and the Euclidean propagator for the positive temperature case takes the following form: $\displaystyle S(k_n)=(k_n^2+m^2)^{-\gamma}$, with $k_n^2=\omega_n^2+k^2$, $\omega_n=2n\pi/\beta$, and $1/\beta=T$  is the temperature. The free energy corresponds to the type I fractional Klein-Gordon field equals to that for ordinary Klein-Gordon field multiply by a factor $\gamma$. When $D=3$, the high temperature limit of the free energy is\cite{54}~:
$$-\gamma\left(\frac{\pi^2}{90}T^4+\frac{1}{24}m^2T^2-\frac{1}{12\pi}m^3T+\ldots\right).$$
In general, it is very difficult to compute the free energy for the type III fractional Klein-Gordon field. The only case  it has been computed is when $D=0$\cite{n28}~. After renormalization, the free energy is given by
\begin{align*}
\mathcal{F} =&\frac{\gamma}{2\beta}\Biggl\{ \log
m^{2\alpha}+\alpha \log\beta^2-2 \sum_{\substack{l\in\mathbb{N}\\l\neq
\frac{1}{2\alpha}}}\frac{(-1)^l}{l}
 \left(\frac{\beta m}{2\pi}\right)^{2\alpha l}\zeta_R(2\alpha
l)\\&+  \omega_{\alpha }
(-1)^{\frac{1}{2\alpha}}\frac{\beta m}{\pi}
\left[\alpha\left( \log \left(\frac{2\pi}{\beta
m }\right)^2+2\psi(1)\right)-\psi\left(\frac{1}{2\alpha}
\right)+\psi(1)\right]\Biggr\}.
\end{align*}Here $\displaystyle \zeta_R(s)=\sum_{n=1}^{\infty}n^{-s}$ is the Riemann zeta function and $\omega_{\alpha}=1$ if and only if $1/(2\alpha)$ is   an integer.
Otherwise $\omega_{\alpha}=0$. The dependence of the dimensionless free energy $\mathcal{F}/m$ on $\beta m$ and $\alpha$ are shown in Fig. \ref{f2} and Fig. \ref{f1}.

Casimir energy due to massless field between two parallel plates subject to fractional Neumann boundary conditions has been computed based on the zeta function regularization method \cite{n18}~. Let us recall briefly the basic steps involved in this regularization procedure which can be summarized in three steps. In the case of scalar massless fractional Klein-Gordon field, they are (i) Evaluation of eigenvalues $\lambda$  of the fractional D$^{\prime}$Alembertian operator $(-\Delta)^{\alpha}$  with certain boundary conditions, thus leading to spectral zeta function $\displaystyle \zeta_{(-\Delta)^{\alpha}}(s)=\sum_{\lambda}\lambda^{-s}$; (ii) Analytic continuation of the zeta function to a meromorphic function of the whole complex plane; and (iii) Calculation of $\det (-\Delta)^{\alpha}$  in terms of  $\zeta_{(-\Delta)^{\alpha}}(s)$ using the relation $\displaystyle\det (-\Delta)^{\alpha} =\exp\left(-\zeta_{(-\Delta)^{\alpha}}'(0)\right)$.

Fractional Neumann boundary conditions were introduced to study the Casimir effect due to fractional Klein-Gordon field. On two plates separated by a distance $a$, the fractional Neumann boundary conditions are:
\begin{equation*}
\begin{split}
\left.\frac{\pa^{\mu}}{\pa x_D^{\mu}}\phi\left(t,\bar{x},x_D\right)\right|_{x_D=0}=0,\\
\left.\frac{\pa^{\nu}}{\pa x_D^{\nu}}\phi\left(t,\bar{x},x_D\right)\right|_{x_D=a}=0,
\end{split}
\end{equation*}
with  $\bar{x}\in\mathbb{R}^{D-1}$, $x_D\in \mathbb{R}$ and $\mu,\nu\in [0,1]$, and the fractional derivative is defined in terms of its Fourier transform:
\begin{align*}
\frac{\pa^{\mu}f(x)}{\pa x^{\mu}}=\frac{1}{2\pi}\int_{-\infty}^{\infty} dk (ik)^{\mu}e^{ikx}\hat{f}(-k),
\end{align*}
where $(\pm ik)^{\alpha}=|k|^{\alpha}e^{\pm i\alpha\pi/2}\text{sign}(k)$.

Results on free energy density and pressure due to fractional massless field confined to parallel plates subject to the following fractional Neumann boundary conditions have been obtained\cite{58}~. These include (a) General fractional Neumann boundary conditions with $\mu\neq \nu$  and  $(\mu,\nu)\neq (0,1)$   or $(1,0)$; (b) $\mu=\nu\neq 0, 1$; (c) Dirichlet boundary conditions $\mu=\nu=0$; (d) Neumann boundary conditions $\mu=\nu=1$; and (e) Boyer boundary conditions   $(\mu,\nu)= (0,1)$   or $(1,0)$. Low and high temperature limits of the Casimir energy and pressure have been calculated for the above cases. It is interesting to note that there exists a transition value in the difference in the orders of the fractional Neumann conditions for the two plates, that is $|\mu-\nu|$, for which the Casimir force changes from attractive to repulsive, or vice versa.

\section{Topological Symmetry Breaking of Self-interacting Fractional Klein-Gordon Field}\label{sec4}
Quartic self-interacting fractional Klein-Gordon scalar field in a toroidal spacetime and the problem of topological mass generation and symmetry breaking has been investigated \cite{n1}~. The Lagrangian is
\begin{equation*}
L=-\frac{1}{2}\phi(t,x)\Lambda(\pa)\phi(t,x)-\frac{\lambda}{4!}\phi^4(t,x),
\end{equation*}
where $\Lambda(\pa)=(-\Delta+m^2)^{\gamma}$	    	 is the  type I fractional Klein-Gordon operator. Based on the Epstein zeta function regularization method, the one-loop effective potential for both the massive and massless $\phi^4$ theory in the toroidal spacetime $T^p\times \mathbb{R}^q$ is derived in terms of power series of $\lambda\tilde{\varphi}^2$, where  $\tilde{\varphi}$ is the constant classical background field. Renormalization of these quantities is carried out, and the results for the renormalized mass are obtained explicitly.
In the massive case, the renormalized mass $m_{\text{ren}}$ is related to the bare mass $m$ by
\begin{align*}
m_{ \text{ren}}^{2\gamma}=&m^{2\gamma}+\frac{\lambda\pi^{\gamma}m^{\frac{d}{2}-\gamma}}{(2\pi)^{\frac{d}{2}+\gamma}\Gamma(\gamma)}
\times\\ &\sum_{\mathbf{k}\in \mathbb{Z}^p\setminus\{
\mathbf{0}\}}\left(\sum_{i=1}^p
\left[L_ik_i\right]^2\right)^{-\frac{d-2\gamma}{4}}
K_{\frac{d-2\gamma}{2}} \left(m\sqrt{\sum_{i=1}^p \left[L_i
k_i\right]^2}\right).\nonumber
\end{align*}Here   $d=D+1=p+q$, $L_1,\ldots, L_p$ are the compactification lengths of the torus $T^p$. In the massless $m=0$ case, the theory is nonrenormalizable when $\alpha=d/2$. When $\alpha\neq d/2$,
the renormalized  topologically generated mass $m_{\text{ren}}$ is

\vspace{0.2cm}\noindent $\bullet$\;\; If $p=0$, then
$m_{ \text{ren}}^{2\gamma}=0$;

\vspace{0.2cm}\noindent $\bullet$\;\; If $p\geq 1$, and

 \vspace{0.2cm}
$\bullet$\;\; if $\gamma \neq \frac{q}{2}$, then
\begin{align*}
m_{ \text{ren}}^{2\gamma}=&\frac{\lambda}{\Gamma(\gamma)}\frac{1}{2^{q+1}\pi^{\frac{q}{2}}\left[\prod_{i=1}^p
L_i\right]}\Gamma\left(\gamma-\frac{q}{2}\right)Z_{E,p}\left(\gamma-\frac{q}{2};
\frac{2\pi}{L_1},\ldots,\frac{2\pi}{L_p}\right)\\
=&\frac{\lambda}{\Gamma(\gamma)}\frac{1}{2^{2\gamma+1}\pi^{\frac{d}{2}}}\Gamma\left(\frac{d}{2}-\gamma\right)
Z_{E,p}\left( \frac{d}{2}-\gamma; L_1, \ldots, L_p\right).
\end{align*}

\vspace{0.2cm} $\bullet$\;\; if $\gamma = \frac{q}{2}$, then
\begin{align*}
m_{ \text{ren}}^{2\gamma}=&\frac{\lambda}{\Gamma(\gamma+1)}\frac{1}{2^{q+1}\pi^{\frac{q}{2}}\left[\prod_{i=1}^p
L_i\right]}\\&\times\left\{ 1+\gamma\Bigl[\psi(\gamma)-\psi(1)\Bigr]+\gamma
Z_{E,p}'\left( 0 ; \frac{2\pi}{L_1}, \ldots,
\frac{2\pi}{L_p}\right)\right\}.
\end{align*}
Here $Z_{E,p}(s; a_1, \ldots,
a_p)$ is the zeta function
\begin{align*}
Z_{E,p}(s; a_1, \ldots,
a_p)=\sum_{\mathbf{k}\in\mathbb{Z}^p}\left(\sum_{i=1}^p
[a_ik_i]^2\right)^{-s}.
\end{align*}
From these, conditions for symmetry breaking are derived analytically, which show that there is no symmetry breaking in the massive case, however, for the massless case, symmetry breaking occurs provided $0<d-2\gamma\leq p$ for a fixed compactified dimension $p\leq 9$. In the case $p\geq 10$, symmetry breaking can only appear for $d-2\gamma$   in a proper subset of $(0,p]$.

\begin{figure}\centering \epsfxsize=.6\linewidth
\epsffile{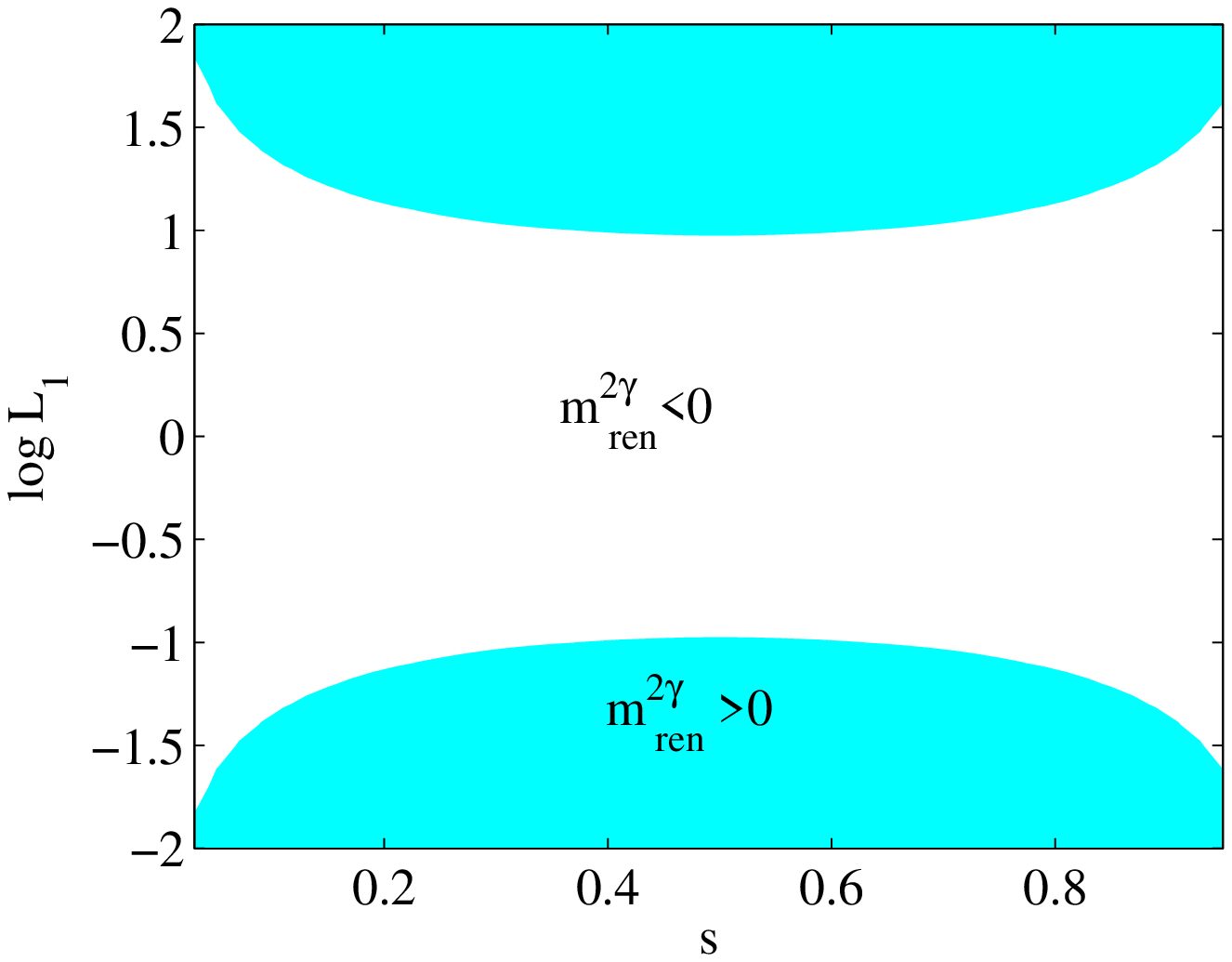} \caption{  \label{f7}The   regions where
$m_{\text{ren}}^{2\gamma}>0$ and $m_{\text{ren}}^{2\gamma}<0$ for $p=2$
and $ L_1L_2=1$.  Here
$s=\frac{d}{2}-\gamma$.}\centering\epsfxsize=.6\linewidth
\epsffile{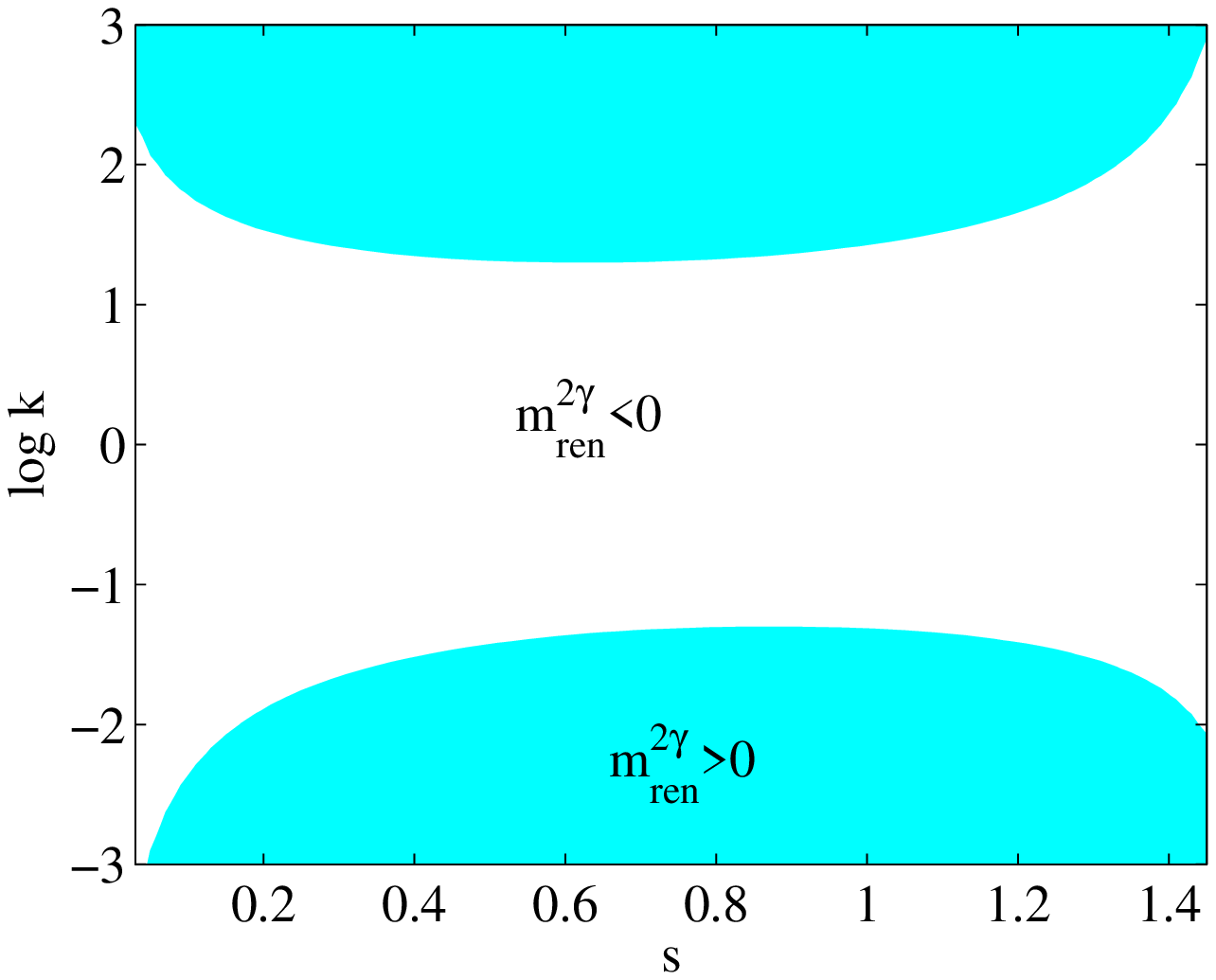}\caption{ \label{f8} The regions where
$m_{ \text{ren}}^{2\gamma}>0$ and $m_{ \text{ren}}^{2\gamma}<0$
for $p=3$, $ L_1L_2L_3=1$, $L_1:L_2:L_3=k:1:1$. Here
$s=\frac{d}{2}-\gamma$.}\end{figure}

\begin{figure}\centering \epsfxsize=0.6 \linewidth\epsffile{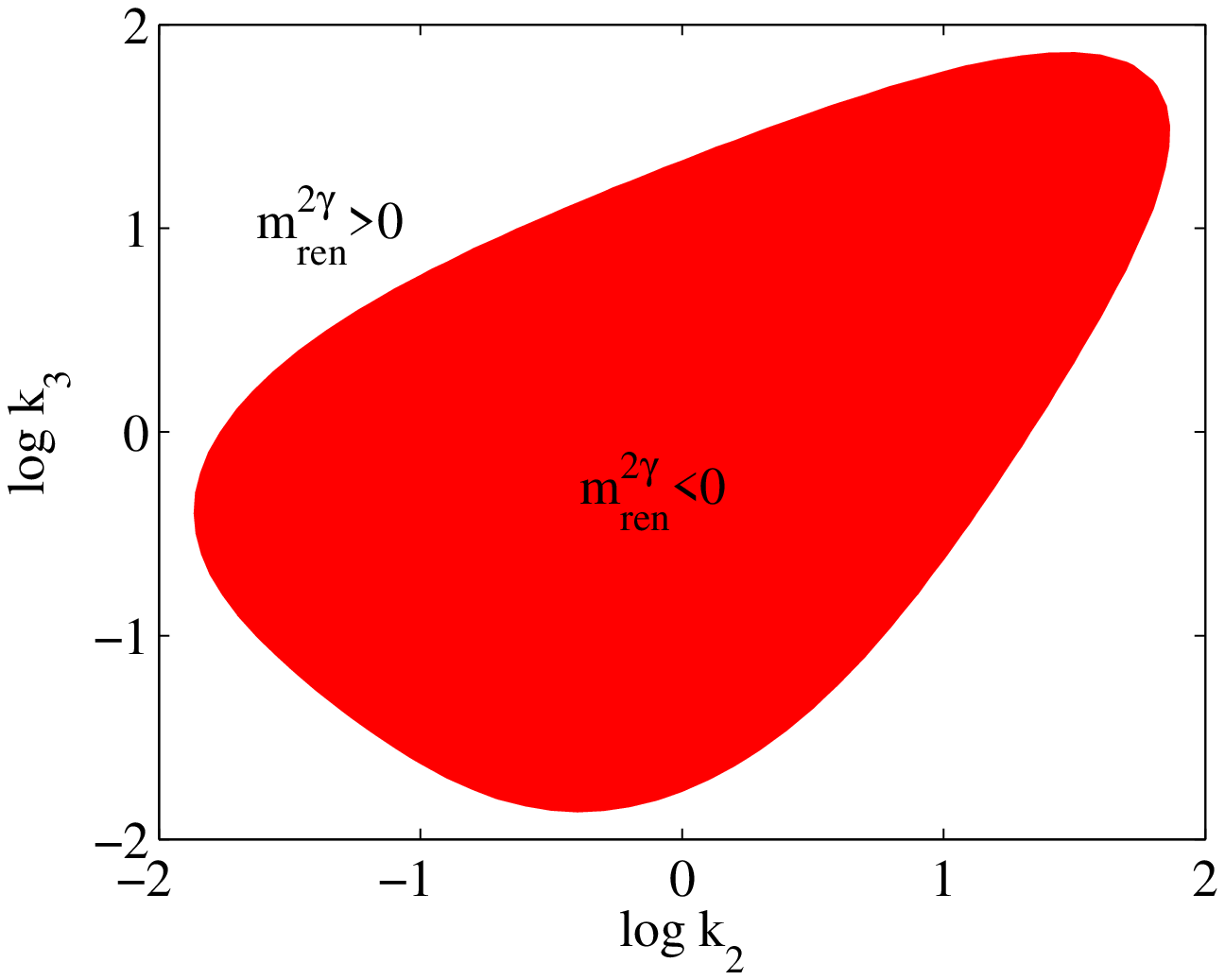} \caption{\label{f9}The region
where $m_{\text{ren}}^{2\gamma}>0$ and
$m_{\text{ren}}^{2\gamma}<0$ for $p=3$,
 $ L_1L_2L_3=1$,
$L_1:L_2:L_3=1:k_2:k_3$ and  $\frac{d}{2}-\gamma=0.1$.}
 \centering\epsfxsize=.6 \linewidth\epsffile{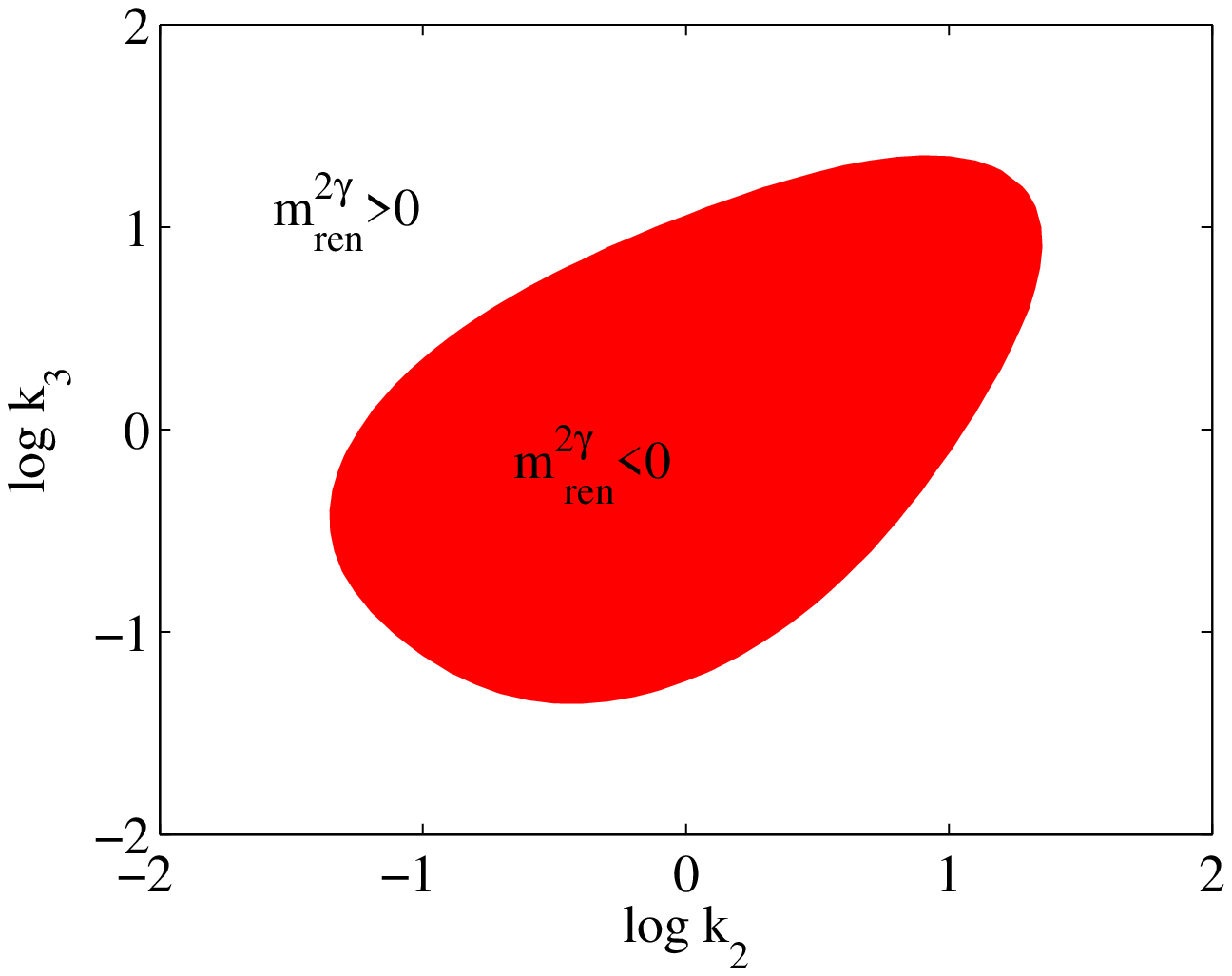} \caption{\label{f10}The regions
where $m_{\text{ren}}^{2\gamma}>0$ and
$m_{\text{ren}}^{2\gamma}<0$ for $p=3$,
 $ L_1L_2L_3=1$, $L_1:L_2:L_3=1:k_2:k_3$and  $\frac{d}{2}-\gamma=0.3$.
 }\centering\epsfxsize=0.6 \linewidth
\epsffile{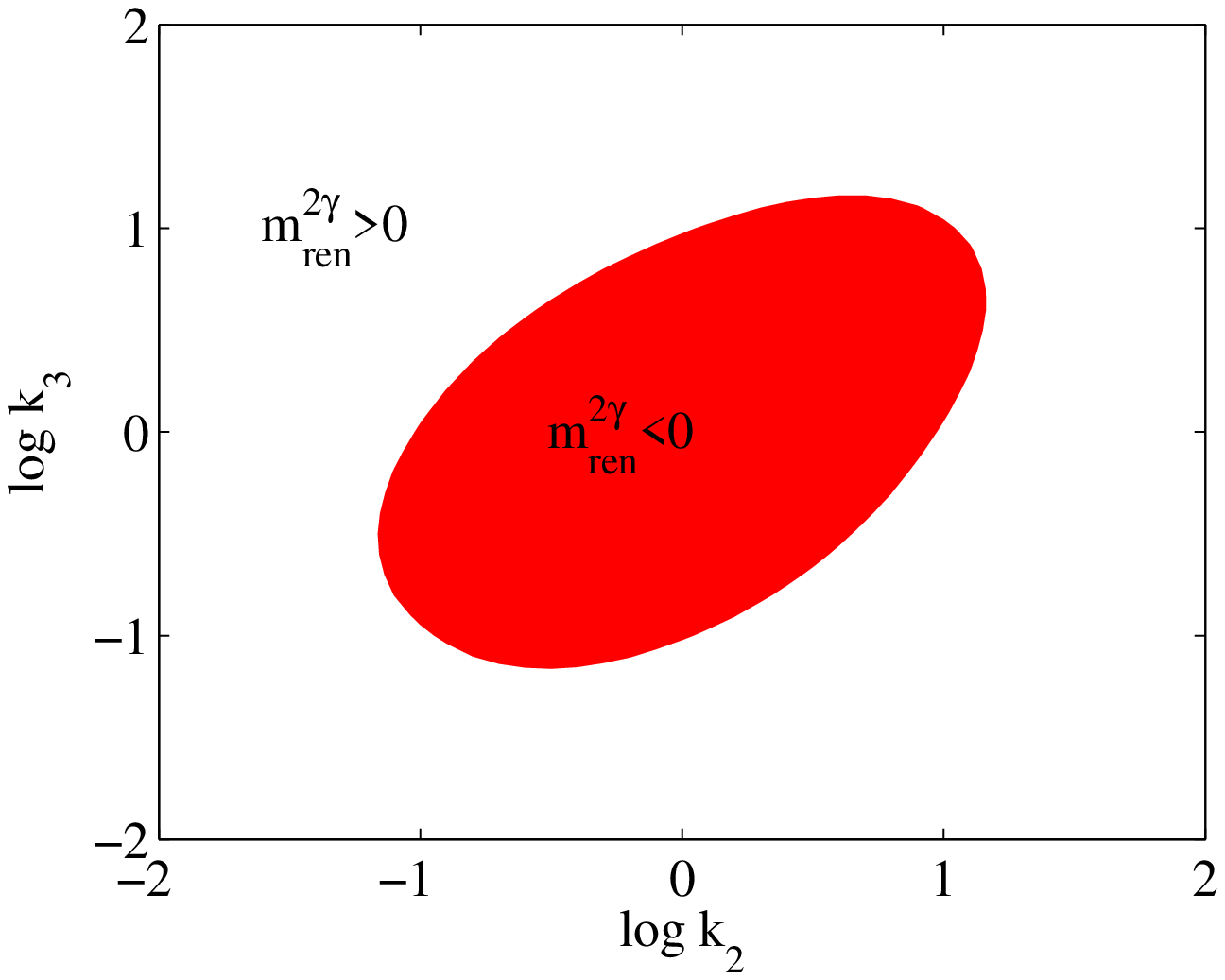} \caption{\label{f11} The regions
where $m_{\text{ren}}^{2\gamma}>0$ and
$m_{\text{ren}}^{2\gamma}<0$ for $p=3$,
 $ L_1L_2L_3=1$,
$L_1:L_2:L_3=1:k_2:k_3$ and  $\frac{d}{2}-\gamma=0.6$. }

\end{figure}

 In Fig. \ref{f7}, Fig. \ref{f8}, Fig. \ref{f9}, Fig. \ref{f10} and Fig. \ref{f11},    the regions where the
renormalized topologically generated mass $m_{ \text{ren}}^{2\gamma}$  are positive and negative respectively are shown.

\section{Casimir Piston Associated with Massless Fractional Klein-Gordon Field}\label{sec5}

In the calculations of Casimir force based on the conventional methods for a confined geometry such as a rectangular cavity, one usually ignores the nontrivial contribution of vacuum energy outside the cavity. A finite result is obtained by discarding the surface divergent terms which depend on the dimensions and geometry under the pretext of some regularization methods \cite{63,64,65,66}~. However, there are studies which show that renormalization of physical parameters of the theory fails to remove such surface divergence \cite{67,68}~.

In 2004, Cavalcanti \cite{69} introduced a new geometric setup, namely a two-dimensional rectangular piston and showed that for such a geometrical configuration it is possible to obtain an unambiguous finite Casimir force despite the presence of the divergence in the Casimir energy. He demonstrated that in the case of a massless scalar field with Dirichlet boundary conditions, a finite attractive Casimir force without the surface divergence is obtained due to the cancelation of the divergent part of the Casimir force acting in the two regions separated by the piston. This result leads to a surge in the studying of the Casimir force on pistons with various geometric configurations and boundary conditions \cite{n2,n3,n4,n5,n6,n7,n21,n22,n23,n14,n15,n16,n24,n17,n18,n19,n20}~.  Casimir force on a rectangular piston in the spacetime with extra compactified dimensions has also been studied \cite{n8,n9,n10,n11,n12,n13}~. It has  been shown that the Casimir force between two (nonmagnetic) dielectric bodies which are related by reflection is always attractive \cite{n25,n26}~. The attractive nature of the Casimir force may result in undesirable effects stiction in nano devices. As a result, it is desirable to search for circumstances under which the Casimir force can become less attractive, or even repulsive. Results obtained by Barton \cite{n21} showed that for a thin piston with weakly reflecting dielectrics, the nature of Casimir force is separation-dependent. The force is attractive at small separations, but it becomes repulsive when the separation increases. For Casimir effect on an $n$-dimensional ($n$ = 1, 2 or 3) rectangular piston with mixed boundary conditions, namely with one surface having Dirichlet condition, and the other with Neumann condition, the resulting force is repulsive \cite{n22}. Other possible ways of obtaining a repulsive Casimir force have also been discussed \cite{n23,n24,n18,n19,n12,n27}~. In particular,   the Casimir force of fractional Klein-Gordon field acting on a piston with fractional boundary conditions can be repulsive \cite{n18, n27}~.

Finite temperature Casimir force acting on a $(D+1)$-dimensional   piston in a $(D+1)$-dimensional semi-infinite Dirichlet cylinder  due to a type I fractional massless Klein-Gordon field at positive temperature has been calculated \cite{n18,n27}~. For such a system, when the   fractional Neumann condition of order $\mu$  is imposed on the piston,
the Casimir force acting on the piston is given by
\begin{equation*} F = -2\gamma T\text{Re}\;\sum_{j }\sum_{l=-\infty}^{\infty}\frac{e^{i\pi \mu}\sqrt{m_{j }^2+(2\pi l T)^2}}{\exp\left(2a\sqrt{m_{j }^2+(2\pi l T)^2}\right)-e^{i\pi  \mu}},\end{equation*}
where $a$ is the distance between the piston and the opposite wall; $\{m_j\}$ is the set of eigenvalues of the Laplacian operator with Dirichlet boundary conditions on the cross section of the cylinder.
\begin{figure}\centering \epsfxsize=.7\linewidth
\epsffile{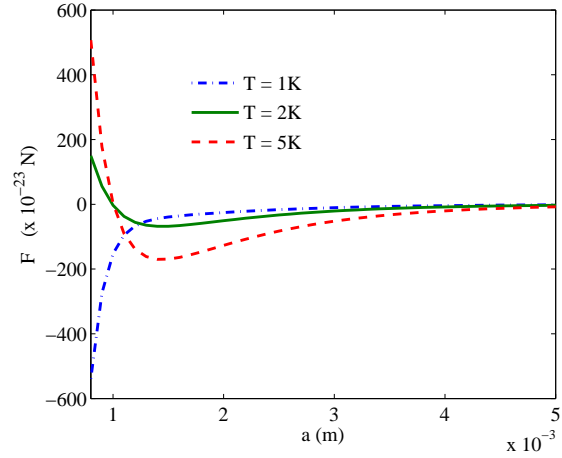}\caption{\label{f3} The Casimir force $F$ as a function of $a$ when $d=3$, $L_2=L_3=0.01$m, $\mu=0.47$, and $T= 1$K, 2K and 5K respectively. }\end{figure}
\begin{figure}\centering \epsfxsize=.7\linewidth
\epsffile{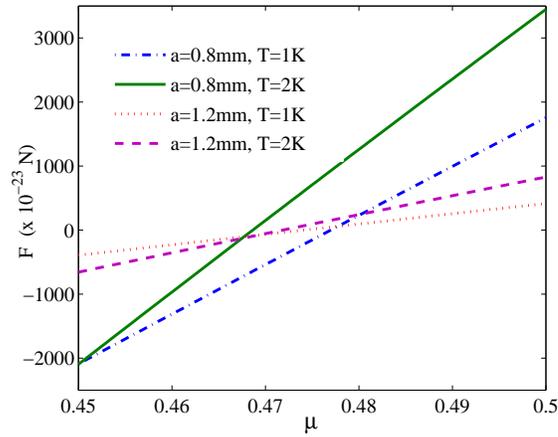}\caption{\label{f4} The Casimir force $F$ as a function of $\mu$ when $d=3$, $L_2=L_3=0.01$m for various values of $(a, T)$.}\end{figure}

\begin{figure}\centering \epsfxsize=.7\linewidth
\epsffile{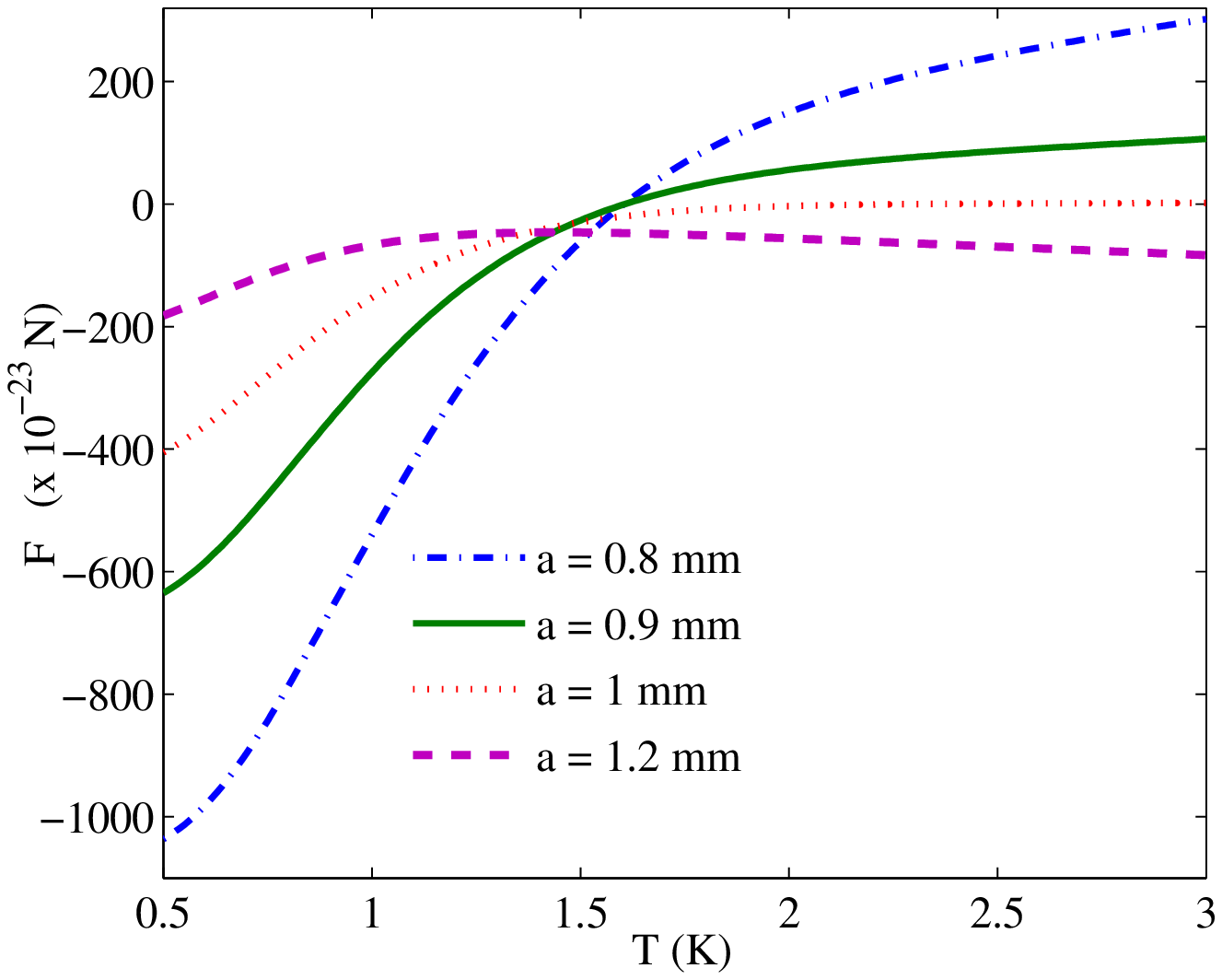} \caption{\label{f5} The Casimir force $F$ as a function of $T$ when $d=3$, $L_2=L_3=0.01$m, $\mu=0.47$  and $a= 0.8$mm, 0.9mm, 1mm and 1.2mm respectively.}\end{figure}
\begin{figure} \centering \epsfxsize=.7\linewidth
\epsffile{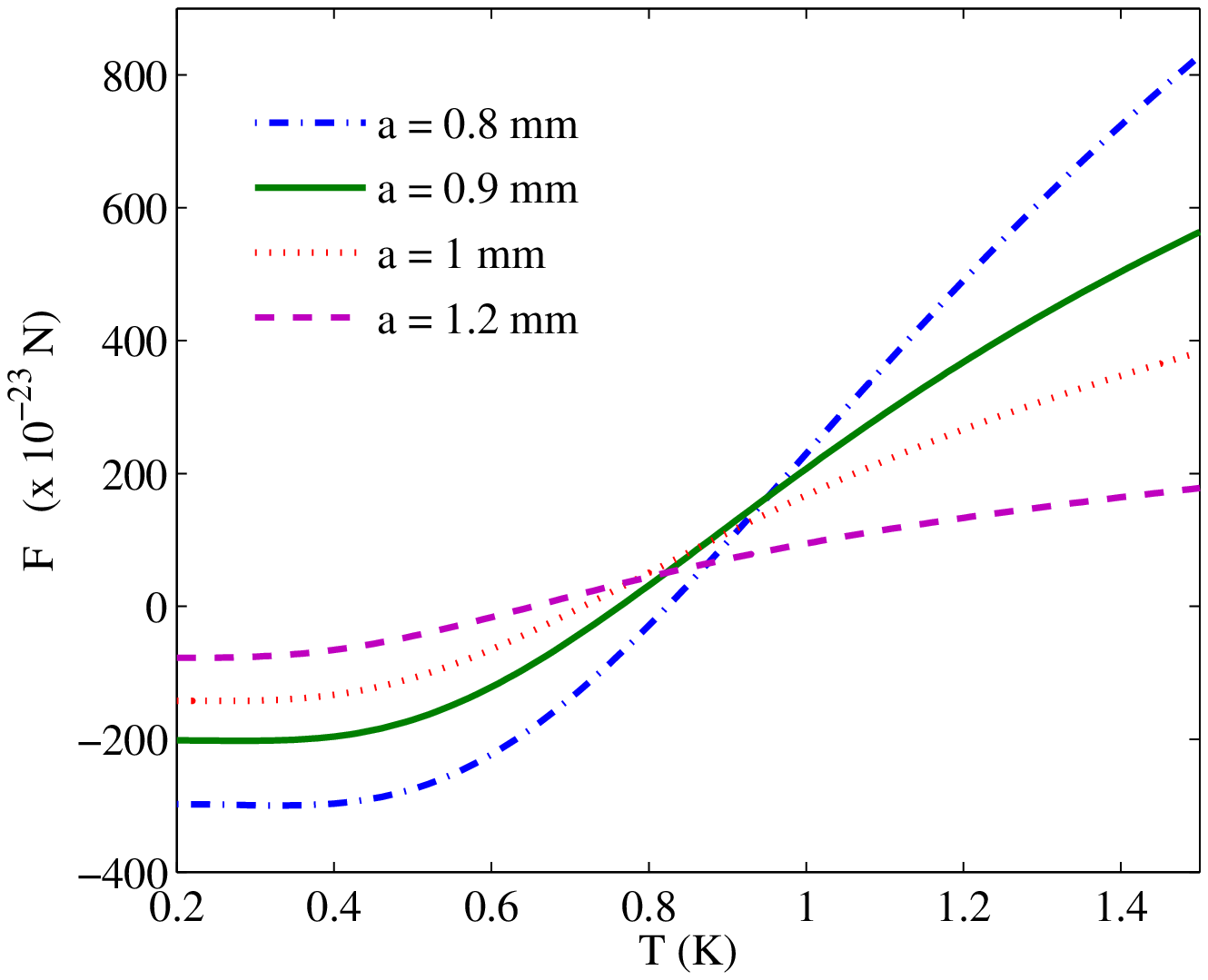}\caption{\label{f6} The Casimir force $F$ as a function of $T$ when $d=3$, $L_2=L_3=0.01$m,  $\mu=0.48$   and $a= 0.8$mm, 0.9mm, 1mm and 1.2mm respectively.}\end{figure}
This result has to be divided by two if $\mu=0$ (Dirichlet) or $\mu=1$ (Neumann).
From this, we find that for $\mu>1/2$, the Casimir force acting on the piston is always repulsive, and the magnitude of the force decreases with $a$. On the other hand, when $\mu<1/2$, the Casimir force can change from repulsive to attractive with increasing  $a$. For certain values of $\mu$  less than but close to $1/2$, the Casimir force can be attractive or repulsive depending on the aspect ratio of the cavity and the temperature. Increasing temperature can change the Casimir force from attractive to repulsive. In the case $D=3$ and  the cross section of the cylinder is the rectangle box $[0,L_2]\times[0,L_3]$, the dependence of the Casimir force on   various quantities  are shown graphically in Fig. \ref{f3}, Fig. \ref{f4}, Fig. \ref{f5} and Fig. \ref{f6}.

\section{Casimir Piston Associated with Massive Fractional Klein-Gordon Field}\label{sec6}

\begin{figure}[h]\centering \epsfxsize=.6\linewidth
\epsffile{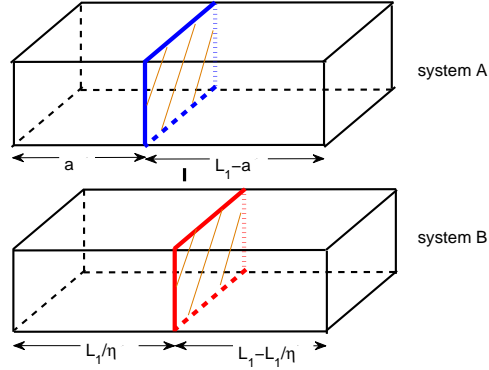}\caption{\label{f12} Regularizing of Casimir energy.}\end{figure}

In this section, we present some new results on the Casimir effect of a massive fractional Klein-Gordon field in the piston scenario.
Consider the type III fractional Klein Gordon  field $\phi(t, \mathbf{x})$  satisfying the  equation:
\begin{equation*}
\left[(-\Delta)^{\alpha}+m^{2\alpha} \right]^{\gamma}\phi(t, \mathbf{x}) =0.
\end{equation*}
We are interested in the finite temperature Casimir effect of the fractional Klein--Gordon field $\phi(t, \mathbf{x})$ on a piston   inside a  semi-infinite rectangular cylinder $[0,\infty]\times[0,L_2]\times\ldots\times[0,L_D]$, with Dirichlet boundary conditions    on the walls of the cylinder and fractional Neumann condition
of order $\mu$ on the piston. The rectangular piston is located at $x_1=a$. To regularize the Casimir energy, first assume that the system is enclosed in the rectangular box $[0,L_1]\times\ldots\times[0,L_D]$. Take  another piston system  where   the piston is located at $x_1=L_1/\eta$, for a constant $\eta>1$.  The regularized Casimir energy  is defined as the limit (see Fig. \ref{f12})
\begin{equation*}
E_{\text{Cas}} =\lim_{L_1\rightarrow \infty}\left(E_{\text{Cas}}^{\text{A}}-E_{\text{Cas}}^{\text{B}}\right),
\end{equation*}where $E_{\text{Cas}}^{\text{A}}$ and $E_{\text{Cas}}^{\text{B}}$ are respectively the Casimir energies of system with the piston at $x_1=a$ and at $x_1=L_1/\eta$ respectively.
\begin{equation*}
E_{\text{Cas}}^{\text{A}}=E_{\text{Cas}}^{\text{box}}(a)+E_{\text{Cas}}^{\text{box}}(L_1-a),\hspace{1cm}
E_{\text{Cas}}^{\text{B}}=E_{\text{Cas}}^{\text{box}}\left(\frac{L_1}{\eta}\right)+E_{\text{Cas}}^{\text{box}}\left(L_1-\frac{L_1}{\eta}\right).
\end{equation*}Using  zeta regularization, the finite temperature Casimir energy inside the box $[0,a]\times[0,L_2]\times\ldots\times[0,L_D]$ is given by
\begin{equation*}
\begin{split}
E_{\text{Cas}}^{\text{box}}(a)=-\frac{T}{2}\left(\zeta_T'(0;a)+\log[\nu^2]\zeta_T(0;a)\right),
\end{split}
\end{equation*}where $\nu$ is a normalization constant, and $\zeta_T(s;a)$ is the zeta function
\begin{equation*} \begin{split}
\zeta_T(s;a)=&  \sum_{n=-\infty}^{\infty}\sum_{\mathbf{k}_{\perp}\in \mathbb{N}^{D-1}}\sum_{l=-\infty}^{\infty}\left\{\left(\omega_{\mathbf{k}_{\perp},l}^2+\left[\frac{\pi \left(n-\frac{\mu}{2}\right)}{a}\right]^2\right)^{\alpha}+m^{2\alpha}\right\}^{-\gamma s}.\end{split}
\end{equation*} Here $\mathbf{k}_{\perp}=(k_2,\ldots,k_D)$,
$$\omega_{\mathbf{k}_{\perp},l}^2=\omega_{\mathbf{k}_{\perp}}^2+[2\pi l T]^2,\hspace{1cm}\omega_{\mathbf{k}_{\perp}}^2=\left(\frac{\pi k_2}{L_2}\right)^2+\ldots+\left(\frac{\pi k_D}{L_D}\right)^2.$$ When $\mu=0$ or $\mu=1$, the summation over $n$ starts from $n=1$.
Using standard techniques,
\begin{equation*}
\begin{split}
\zeta_T(s;a)=& \frac{2}{\Gamma(\gamma s)} \sum_{n=-\infty}^{\infty}\sum_{\mathbf{k}_{\perp}\in \mathbb{N}^{D-1}}\sum_{l=0}^{\infty}\!'
\int_0^{\infty} t^{\gamma s-1}\\&\hspace{1cm}\times\exp\left(-t\left\{\left(\omega_{\mathbf{k}_{\perp},l}^2+\left[\frac{\pi \left(n-\frac{\mu}{2}\right)}{a}\right]^2\right)^{\alpha}+m^{2\alpha}\right\}\right) dt\\
=&\frac{2 }{\Gamma(\gamma s)}\sum_{j=0}^{\infty}
\frac{(-1)^j}{j!}m^{2\alpha j}\Gamma(\gamma s+j) Z\left(\alpha(\gamma s+j) \right),
\end{split}
\end{equation*}
where
\begin{equation*}
Z(s)=\sum_{n=-\infty}^{\infty}\sum_{\mathbf{k}_{\perp}\in \mathbb{N}^{D-1}}\sum_{l=0}^{\infty}\!'\left(\omega_{\mathbf{k}_{\perp},l}^2+\left[\frac{\pi\left(n-\frac{\mu}{2}\right)}{a}\right]^2 \right)^{-s}.
\end{equation*}
Using the fact that
$$\sum_{n=-\infty}^{\infty}  \exp\left(-t\left[\frac{\pi \left(n-\frac{\mu}{2}\right)}{a}\right]^2\right)=\frac{a}{\sqrt{\pi t}}\sum_{n=-\infty}^{\infty} \exp\left(-\frac{n^2a^2}{t}\right)e^{\pi i n\mu},$$ we find that
\begin{equation*}
\begin{split}
 Z(s)=&\frac{1}{\Gamma(s)}\int_0^{\infty} t^{s-1}\sum_{n=-\infty}^{\infty}\sum_{\mathbf{k}_{\perp}\in \mathbb{N}^{D-1}}\sum_{l=0}^{\infty}\!'\exp\left\{-t\left(\left[\frac{\pi \left(n-\frac{\mu}{2}\right)}{a}\right]^2+\omega_{\mathbf{k}_{\perp},l}^2 \right)\right\}dt\\
 =& \frac{a}{\sqrt{\pi}\Gamma(s)}\int_0^{\infty} t^{s-\frac{3}{2}}\sum_{n=-\infty}^{\infty} \sum_{\mathbf{k}_{\perp}\in \mathbb{N}^{D-1}}\sum_{l=0}^{\infty}\!'e^{\pi i n\mu}\exp\left\{-t \omega_{\mathbf{k}_{\perp},l}^2 -\frac{n^2a^2}{t}\right\}dt\\
 =& \frac{a}{ \sqrt{\pi}\Gamma(s)}\Gamma\left(s-\frac{1}{2}\right)Z_0\left(s-\frac{1}{2}\right)
  \\&+\frac{4a}{\sqrt{\pi}\Gamma(s)}\sum_{n=1}^{\infty}\sum_{\mathbf{k}_{\perp}\in \mathbb{N}^{D-1}}\sum_{l=0}^{\infty}\!'e^{\pi i n\mu}\left(
 \frac{na}{ \omega_{\mathbf{k}_{\perp},l} }\right)^{s-\frac{1}{2}}K_{s-\frac{1}{2}}\left(2na \omega_{\mathbf{k}_{\perp},l} \right),
\end{split}
\end{equation*} where $$Z_0(s)= \sum_{\mathbf{k}_{\perp}\in \mathbb{N}^{D-1}}\sum_{l=0}^{\infty}\!' \omega_{\mathbf{k}_{\perp},l}^{-2s}.$$Therefore,
\begin{equation*}
\begin{split}\zeta_T(s;a)= &aY(s)+\frac{8a }{\sqrt{\pi}\Gamma(\gamma s)}\sum_{j=0}^{\infty}
\frac{(-1)^j}{j!}m^{2\alpha j}\frac{\Gamma(\gamma s+j)}{\Gamma\left(\alpha(\gamma s+j)\right)}
\sum_{n=1}^{\infty}\sum_{\mathbf{k}_{\perp}\in \mathbb{N}^{D-1}}\sum_{l=0}^{\infty}\!'\\
&\hspace{2cm}\times e^{\pi i n\mu} \left(
 \frac{na}{ \omega_{\mathbf{k}_{\perp},l} }\right)^{\alpha(\gamma s+j)-\frac{1}{2}}K_{\alpha(\gamma s+j)-\frac{1}{2}}\left(2na \omega_{\mathbf{k}_{\perp},l} \right),
\end{split}
\end{equation*}
where  $Y(s)$ is independent of $a$. One immediately finds that
$$\zeta_T(0;a)= aY_1(0),$$and
\begin{equation*}
\begin{split}\zeta_T'(0;a)=&aY_1'(0)+\frac{8 \alpha \gamma a}{\sqrt{\pi} }\sum_{j=0}^{\infty}
 (-1)^j \frac{m^{2\alpha j}}{\Gamma\left(\alpha j+1\right)}
\sum_{n=1}^{\infty}\sum_{\mathbf{k}_{\perp}\in \mathbb{N}^{D-1}}\sum_{l=0}^{\infty}\!'e^{\pi i n\mu}\\&\hspace{2cm}\times \left(
 \frac{na}{ \omega_{\mathbf{k}_{\perp},l} }\right)^{\alpha j-\frac{1}{2}}K_{\alpha j-\frac{1}{2}}\left(2na \omega_{\mathbf{k}_{\perp},l} \right).
\end{split}
\end{equation*}
  From this, we find that
the finite temperature Casimir energy in the box $[0,a]\times[0,L_2]\times\ldots\times[0,L_D]$  is given by
\begin{equation*}\begin{split}
E_{\text{Cas}}^{\text{box}}(a)=& aE_1-\frac{4\alpha\gamma a T }{\sqrt{\pi} }\sum_{j=0}^{\infty}
 (-1)^j \frac{m^{2\alpha j}}{\Gamma\left(\alpha j+1\right)}
\sum_{n=1}^{\infty}\sum_{\mathbf{k}_{\perp}\in \mathbb{N}^{D-1}}\sum_{l=0}^{\infty}\!'e^{\pi i n\mu}\\&\hspace{2cm}\times\left(
 \frac{na}{ \omega_{\mathbf{k}_{\perp},l} }\right)^{\alpha j-\frac{1}{2}}K_{\alpha j-\frac{1}{2}}\left(2na \omega_{\mathbf{k}_{\perp},l} \right),
\end{split}\end{equation*}where  the term $E_1$ is independent of $a$. This term will vanish when we take the difference of the energies in system A and system B. In the limit $L_1\rightarrow \infty$, we find that the finite temperature Casimir energy of the piston system is
\begin{equation}\label{eq1_4_1}
\begin{split}E_{\text{Cas}} =&-\frac{4\alpha\gamma a T }{\sqrt{\pi} }\sum_{j=0}^{\infty}
 (-1)^j \frac{m^{2\alpha j}}{\Gamma\left(\alpha j+1\right)}
\sum_{n=1}^{\infty}\sum_{\mathbf{k}_{\perp}\in \mathbb{N}^{D-1}}\sum_{l=0}^{\infty}\!'e^{\pi i n\mu}\\&\hspace{2cm}\times\left(
 \frac{na}{ \omega_{\mathbf{k}_{\perp},l} }\right)^{\alpha j-\frac{1}{2}}K_{\alpha j-\frac{1}{2}}\left(2na \omega_{\mathbf{k}_{\perp},l} \right).
 \end{split}
\end{equation}For $\mu=0$ or $\mu=1$, this result should be divided by two.
Eq. \eqref{eq1_4_1} is the high temperature expansion of the Casimir energy. It shows that the high temperature leading term is given by the sum of the terms with $l=0$, i.e., when $aT\gg 1$,
\begin{equation*}
\begin{split}
E_{\text{Cas}} \sim&-\frac{2\alpha\gamma a T }{\sqrt{\pi} }\sum_{j=0}^{\infty}
 (-1)^j \frac{m^{2\alpha j}}{\Gamma\left(\alpha j+1\right)}
\sum_{n=1}^{\infty}\sum_{\mathbf{k}_{\perp}\in \mathbb{N}^{D-1}} e^{\pi i n\mu}\\&\hspace{2cm}\times\left(
 \frac{na}{ \omega_{\mathbf{k}_{\perp}} }\right)^{\alpha j-\frac{1}{2}}K_{\alpha j-\frac{1}{2}}\left(2na \omega_{\mathbf{k}_{\perp}} \right).
\end{split}
\end{equation*}
The zero temperature Casimir energy is obtained by taking the limit $T\rightarrow 0$ of \eqref{eq1_4_1}, which gives
\begin{equation*}
\begin{split}E_{\text{Cas}}^{  T=0}=&-\frac{2\alpha\gamma a }{\pi^{\frac{3}{2}} }\sum_{j=0}^{\infty}
 (-1)^j \frac{m^{2\alpha j}}{\Gamma\left(\alpha j +1\right)}
\sum_{n=1}^{\infty}\sum_{\mathbf{k}_{\perp}\in \mathbb{N}^{D-1}} e^{\pi i n\mu}\\&\hspace{1.5cm}\times\int_0^{\infty}\left(
 \frac{na}{ \sqrt{\omega_{\mathbf{k}_{\perp} }^2+u^2} }\right)^{\alpha j-\frac{1}{2}}K_{\alpha j-\frac{1}{2}}\left(2na \sqrt{\omega_{\mathbf{k}_{\perp}}^2+u^2} \right)du\\
 =&-\frac{\alpha\gamma a }{\pi }\sum_{j=0}^{\infty}
 (-1)^j \frac{m^{2\alpha j}}{\Gamma\left(\alpha j+1\right)}
\sum_{n=1}^{\infty}\sum_{\mathbf{k}_{\perp}\in \mathbb{N}^{D-1}}e^{\pi i n\mu}\left(\frac{na}{ \omega_{\mathbf{k}_{\perp} } }\right)^{\alpha j-1}\\&\hspace{6cm}\times K_{\alpha j-1}\left(2na \omega_{\mathbf{k}_{\perp} } \right) .
 \end{split}
\end{equation*}
The Casimir force acting on the piston is given by
\begin{equation*}
\begin{split}F_{\text{Cas}}^{\parallel}=&-\frac{\pa E_{\text{Cas}}^{\parallel}}{\pa a}\\=&-\frac{ 8\alpha\gamma  T }{\sqrt{\pi} }\sum_{j=0}^{\infty}
 (-1)^j \frac{m^{2\alpha j}}{\Gamma\left(\alpha j+1\right)}
\sum_{n=1}^{\infty}\sum_{\mathbf{k}_{\perp}\in \mathbb{N}^{D-1}}\sum_{l=0}^{\infty}\!'e^{\pi i n\mu}\\&\hspace{4cm}\times
 \frac{(na)^{\alpha j+\frac{1}{2}}}{ \omega_{\mathbf{k}_{\perp},l}^{\alpha j-\frac{3}{2}} } K_{\alpha j+\frac{1}{2}}\left(2na \omega_{\mathbf{k}_{\perp},l} \right)
\\& +\frac{8\alpha\gamma   T }{\sqrt{\pi} }\sum_{j=1}^{\infty}
 (-1)^j \frac{m^{2\alpha j}}{\Gamma\left(\alpha j \right)}
\sum_{n=1}^{\infty}\sum_{\mathbf{k}_{\perp}\in \mathbb{N}^{D-1}}\sum_{l=0}^{\infty}\!'e^{\pi i n\mu}\\&\hspace{4cm}\times\left(
 \frac{na}{ \omega_{\mathbf{k}_{\perp},l} }\right)^{\alpha j-\frac{1}{2}}K_{\alpha j-\frac{1}{2}}\left(2na \omega_{\mathbf{k}_{\perp},l} \right).
 \end{split}
\end{equation*}In the limit $T\rightarrow 0$, we find that the zero temperature Casimir force acting on the piston is given by
\begin{equation*}
\begin{split}F_{\text{Cas}}^{\parallel, T=0}= &-\frac{2\alpha \gamma  }{ \pi }\sum_{j=0}^{\infty}
 (-1)^j \frac{m^{2\alpha j}}{\Gamma\left(\alpha j+1\right)}
\sum_{n=1}^{\infty}\sum_{\mathbf{k}_{\perp}\in \mathbb{N}^{D-1}} e^{\pi i n\mu}\\&\hspace{4cm}\times\frac{(na)^{\alpha j}}{ \omega_{\mathbf{k}_{\perp} }^{\alpha j-2} } K_{\alpha j }\left(2na \omega_{\mathbf{k}_{\perp} } \right)\\&+\frac{\alpha\gamma   }{ \pi }\sum_{j=0}^{\infty}
 (-1)^j (2\alpha j-1)\frac{m^{2\alpha j}}{\Gamma\left(\alpha j+1\right)}
\sum_{n=1}^{\infty}\sum_{\mathbf{k}_{\perp}\in \mathbb{N}^{D-1}} e^{\pi i n\mu}\\&\hspace{4cm}\times\left(\frac{na}{ \omega_{\mathbf{k}_{\perp} } }\right)^{\alpha j-1}K_{\alpha j-1}\left(2na \omega_{\mathbf{k}_{\perp} } \right).
 \end{split}
\end{equation*}
For a type III massive Klein-Gordon field, we can only expand the Casimir energy and Casimir force as a series in $m^{2\alpha}$. It is not easy to determine the sign of the Casimir force from these series expressions.

\section{Conclusion}\label{sec7}
Justifications on the need to study fractional Klein-Gordon fields are given. A brief summary of the results on fractional Klein-Gordon fields mainly by the author and coworkers are given. We also present new results on Casimir piston associated with fractional Klein-Gordon massive field.

	Finally, we would like to suggest some other possible directions for further work. So far we have considered fractional Klein-Gordon field with unique fractional exponent. Even though type III fractional Klein-Gordon field parametrized by two indices provides more flexibility for modeling, it is still inadequate to represent quantum field in spacetime with dimensions that vary with spacetime scales. In view of the possible variable spacetime dimensions, one may have to deal with fractional quantum fields of variable fractional order, with   replaced by the function  . Fractional stochastic processes of variable order such as multifractional Brownian motion \cite{a ,b}, multifractional L$\acute{\text{e}}$vy motion \cite{c,d}, multifractional Riesz-Bessel process \cite{e} and multifractional Ornstein-Uhlenbeck process \cite{f} have been studied. Variable order one-dimensional Klein-Gordon massive field is given by multifractional Ornstein-Uhlenbeck process, which can be easily extended to higher dimensions to yield multifractional Klein-Gordon field. On the other hand, fractional Klein-Gordon field can also be obtained as a special case of multifractional Riesz-Bessel field. In addition, it may be necessary to impose fractional Neumann boundary conditions of variable fractional order when dealing with multifractional Klein-Gordon field. It should be emphasized that calculations of physical quantities of Casimir effect due to fractional Klein-Gordon field of variable order are expected to be mathematically  difficult, and new techniques will be required.

	Another direction of generalization is to consider the above results in non-flat space, which requires fractional itegro-differential operators in curved spacetime. One particular case of recent interest will be to study fractional Klein-Gordon field in de Sitter spacetime. Stochastic quantization for ordinary Klein-Gordon field in de Sitter spacetime has been carried out \cite{g}. It will be interesting to extend it to the fractional case. Path integral approach to fractional Klein-Gordon field is another area that has so far not been seriously treated.

\end{document}